# Note on the nature of the transition between a system in an equilibrium state and a system in a non-equilibrium state (and vice-versa)

E. G. D. Cohen[1,2, a)] and R. L. Merlino[2, b)]

[1]*The Rockefeller University, 1230 York Avenue, New York, NY 10065*

[2]*The University of Iowa, Iowa City, IA 52242*

[a)] egdc@mail.rockefeller.edu
[b)] robert-merlino@uiowa.edu

August 8, 2013

**Abstract**

The transition from a non-equilibrium state to an equilibrium state is characterized not only by the disappearance of the entropy production, but mainly by the disappearance of the organized currents, due to the gradients present in a non-equilibrium system. Their disappearance is necessary to obtain maximum entropy in the equilibrium state.

In this note we consider the transition from an equilibrium system to a non-equilibrium system, and *vice versa*. While in an equilibrium system there is entropy, a measure of the disorder of the particles in the system, in a non-equilibrium system there is no entropy and only entropy production due to the dissipative processes occurring in non-equilibrium systems, such as e.g., viscous friction, thermal conduction, and particle diffusion. Because of the Second Law of Thermodynamics, these processes are necessarily accompanied by a dissipation of energy in the form of a non-zero entropy production.

Here we will consider fluids near thermal equilibrium, which are in a local equilibrium state. Then local thermodynamic quantities can be defined: a local number density, n($\mathbf{r}$, t), a local temperature T($\mathbf{r}$, t), and a local mean velocity $\mathbf{u}$($\mathbf{r}$, t). Here $\mathbf{r}$ denotes a position in the system and t is the time. This state is the basis of Non-Equilibrium or Irreversible Thermodynamics [1].

In [1], when an equilibrium system makes a transition to a non-equilibrium system, only the appearance of a non-zero entropy production, which vanishes in the equilibrium system, is considered. However, this does not take into account that this transition also involves an organization of the non-equilibrium system. When external forces are applied to bring a system from a homogeneous equilibrium state to an inhomogeneous non-equilibrium state, the system first passes through a non-equilibrium *transient* state, and then to a non-equilibrium steady state. In the transient state, organized currents are formed because of the gradients of the local thermodynamic quantities, as, e.g., in the case of a heat flow or in a Poiseuille flow [2]. In the non-equilibrium steady state, the fully organized currents formed at the end of the transient state, are maintained until the external forces are removed and the system passes again through a transient state and returns to the equilibrium state.

This organization or ordering in a system in a non-equilibrium state leads to a loss of entropy, which is partly compensated by the non-negative entropy production due to dissipation. However, up till now, this has not been considered in Non-Equilibrium Thermodynamics. Consequently, there *has* to be a new (extra) entropic contribution, which in spite of the disappearance of the entropy production, guarantees a maximum of the entropy in equilibrium. This needed entropic contribution is provided by the organized currents, when they disintegrate in the non-equilibrium transient state and disappear in the equilibrium state, because of the disappearance of the gradients. This will randomize the system and then provide the necessary entropic disorder for the maximum entropy in the equilibrium state [3].

Furthermore, in non-equilibrium thermodynamics as in [1], the entropy production, i.e. the dissipation, has a *minimum* in the non-equilibrium steady-state. However, in our case, because of the gradual increase of the entropy production from the equilibrium state via the non-equilibrium transient state to the non-equilibrium steady-state, a *maximum* of the entropy production will be reached in the non-equilibrium steady-state. Physically it seems impossible, that during the steady gradual organization of the currents, from the initial equilibrium state via the non-equilibrium transient state to the final non-equilibrium steady state, the entropy production would have a maximum in the non-equilibrium transient state. This is because the steadily increasing currents from the equilibrium state to the non-equilibrium steady state imply an accompanying steadily increasing dissipation, i.e., entropy production, as shown in Table 1.

| EQUILIBRIUM STATE | NON-EQUILIBRIUM STATES | |
| --- | --- | --- |
| | **Transient state** | **Steady state** |
| entropy | increasing entropy production | maximum entropy production |
| no currents | organizing currents | organized currents |
| no entropy production | no entropy | no entropy |

**TABLE 1.** Entropy, entropy production, and currents in the equilibrium and non-equilibrium states.

For the opposite transition from the non-equilibrium steady state to the equilibrium state, the organization of the currents – due to the externally induced gradients— functions as a *deus ex machina* for the retrieval of the maximum entropy when the system has reached its final equilibrium state.

In conclusion, the main point of this Note is to point out that the entropy of the particles of a fluid in thermal equilibrium is transformed by the external forces into organized currents of the particles in a non-equilibrium state and *vice-versa*.